# Unleashing Excellence through Inclusion: Navigating the Engagement-Performance Paradox

Nicole M. Radziwill & Morgan C. Benton

## Abstract

People who feel that they do not belong (or their voice is not heard at work) commonly become disengaged, unproductive, and pessimistic. Inclusive work environments aspire to close these gaps to increase employee satisfaction while reducing absenteeism and turnover. But there is always a job to be done, and under time and resource constraints, democratic approaches can result in reduced quality and unacceptable delays. Teams need actionable guidance to incorporate inclusive practices that will directly impact effectiveness. This paper contributes to the literature on quality and performance management by developing a conceptual model of inclusion that directly (and positively) impacts performance, and identifies eight factors that workgroups must address to create and maintain inclusive, high performing environments.

## Introduction

In 2024, sentiment around Diversity, Equity, and Inclusion (DEI) has become more polarized within software engineering and related technical areas and beyond. Politicization of DEI efforts, concerns about discrimination, a loss of faith in the effectiveness and implementation of inclusion programs, and the perception that DEI appears to increase prejudice, not reduce it, has led to a deceleration of DEI initiatives. In some cases, there is even "backfire", a documented phenomenon where "well-intended initiatives result in unintended negative outcomes." (Burnett & Aguinis, 2024)

But results are non-negotiable, performance is always a top priority, and consequently companies still need to cultivate inclusive environments to ensure that investments in labor are effective. This is particularly true in software engineering and related technical work, where collaboration is essential to produce high quality products, and the need to "challenge the dominant 'hacker' cultural stereotype" that can inhibit it. (Fraser & Mancl, 2022) As a result, this paper derives a conceptual model and outlines factors that

When employees bring their authentic best selves to work, absenteeism decreases, turnover (and associated costs) go down, and losses due to "language, communication style, and feedback openness" are minimized. (Morfaki & Morfaki, 2022) While these outcomes help organizations meet quality and performance goals, lack of a shared understanding of what is meant by inclusion can inhibit all benefits. (Ely & Thomas, 2001)

Although an inclusive environment makes it *possible* for employees to feel belonging and well-being, creating an environment where most or all *do* often remains elusive. Individual expectations around what inclusion should feel like are often subjective and unachievable, contributing to inclusion's "theoretically unlimited rhetoric". (Hansen, 2012) Additionally, two inescapable facts of the workplace are that roles are defined to distribute effort, and jobs must be performed to standards. Access to all resources and all information is not a given, and not every person is equally qualified to have a meaningful voice in how a particular task is carried out. Both of these factors impact employee perceptions of inclusion.

Although some inclusive leadership enhances productivity, "task performance decreases when inclusive leadership is from moderate to high levels." Inclusive leaders make decisions more slowly because they democratically incorporate complex emotions, which can lead colleagues to believe that they are less effective as leaders. (Xiaotao et al., 2018) This could reduce their commitment to building an inclusive culture, leading to negative feedback loops that ultimately result in less well-being for all. Similarly, a well established concern of democratic systems is that they can be slow, unadaptive, and unresponsive, which can be incompatible with a "high speed world." (Saward, 2017)

This Engagement-Performance paradox requires careful and intentional consideration as organizations incorporate conduct initiatives to improve inclusivity. Teams need to develop "competencies of inclusion" like setting expectations around what is feasible, designing inclusive practices, and continually improving them in a way that is sensitive to goal achievement. (Pless & Maak, 2004) To enable this, "inclusion by design" can help teams subvert the Engagement-Performance paradox and intentionally strike a balance between inclusion and performance. It is this gap that the current study fills.

## What is Inclusion?

Inclusion is the "**intentional, ongoing effort** to ensure that diverse people with different identities are able to **fully participate** in all aspects of the work of an organization, including leadership positions and decision-making processes." (Tan, 2019) When people feel accepted, and feel able to contribute to and influence their work, there is often less absenteeism and turnover. (Mor Barak et al. 2001) An inclusive environment can "access the whole range of its workforce's skills without interference" (Dobusch, 2014) and is sensitive to power relations, especially those that emerge from historical inequalities.

Despite three decades of research exploring how to make organizations more inclusive, *it is still difficult for most organizations to cultivate an inclusive environment in practice.* This is because inclusion is **relational** (it depends on the people we interact with the most), **dynamic** (people, policies, practices, and roles can change which impact our experiences), and **ineffable** (easily misunderstood or conflated with subjective expectations of the feeling of belonging, which is a consequence of inclusion). Visionary values statements, progressive policies, and commitments to inclusion require continuous infusions of energy to constructively impact day to day life.

Inclusion must be addressed at the institutional level, through values and policies, and the team or workgroup level, through inclusive practices. (Silver, 2015) This means ensuring access to resources, access to processes and information, and access to the interpersonal interactions that enable task completion and organizational learning. (Schleien et al. 2003)

|  | Inclusion in: ||
| --- | --- | --- |
|  | **Team/Workgroup** | **Institution** |
| Access to **Resources** *(Physical Inclusion)* | <ul><li>Procure materials required for team's work</li><li>Procure access and authorization for systems required for team's work</li></ul> | <ul><li>Procure materials to participate in the institution (e.g. employee laptop)</li><li>Procure access and authorization to participate in the institution (e.g.</li></ul> |

| | | corporate login, software licenses) |
|---|---|---|
| Access to **Processes and Information** *(Functional Inclusion)* | <ul><li>Open formal channels for communication and information sharing between team members</li><li>Define inclusive practices around:<ul><li>Preparing for meetings</li><li>Participating in meetings</li><li>Accessing meeting records</li><li>Accessing project artifacts</li></ul></li></ul> | <ul><li>Open formal channels for communication and information sharing organization-wide to:<ul><li>Receive and understand official communications</li><li>Provide feedback when requested, or ad hoc ("suggestion box")</li></ul></li><li>Define inclusive practices around:<ul><li>Accessing processes or information relevant to your role</li><li>Accessing records, artifacts, and data relevant to your role</li></ul></li></ul> |
| Access to **Productive Interactions** that stimulate organizational learning *(Social Inclusion)* | <ul><li>Open informal channels for within-team communication</li><li>Implement procedural justice to enable "voice" about team practices</li><li>Cultivate psychological safety to freely express ideas/give honest feedback</li></ul> | <ul><li>Open channels of communication between all levels of the organization</li><li>Implement procedural justice to enable "voice" re:institutional policies</li><li>Cultivate psychological safety to drive out fear of recrimination/job loss</li></ul> |

**Figure 1.** Inclusive environments ensure physical, functional, and social access at workgroup and institutional levels.

## Inclusion is Relational

Inclusion is **relational**. While a dominant group or individual can *create* conditions for inclusion, *it is up to individuals to engage* in the dialogue of inclusion by accepting the resources, processes and information, and interactions that are made available. For social inclusion to be reciprocated, team members must have psychological safety, that is, be *comfortable* sharing ideas, asking questions, initiating collaboration, and providing feedback without fearing negative consequences or ridicule. (Edmondson & Lei, 2014) In all cases, each individual "has a role to play in the creation of their psychological safety," making inclusion an *ongoing dialogue* rather than something that is provided by one party to another. (Mather, 2020)

Psychological safety is important, but *not enough,* to ensure that environments are inclusive. "People who feel [psychologically safe and] free to speak up may simultaneously believe their voice will not be recognized or responded to" (Kerrissey et al., 2022) so additional steps are needed:

- **Discuss immutable forces.** Teams must openly discuss three "immutable forces" - dominant people or groups, unconscious bias, and within-group stereotyping. (Turnbull, 2016) Confronting the reality that different roles, skills, and accountabilities influence voice, and remaining aware of subconscious influences that cannot be removed, can enhance safety.
- **Establish procedural justice.** Teams need to establish procedures for equitably collecting information for decision making from all parties affected, fielding requests for clarification, and making and appealing decisions. This is "procedural justice," one component of organizational justice as defined by Elovainio et al. (2002).
- **Assess psychological capital.** An individual's past experiences and traumas can inhibit feelings of safety even in environments where others feel very psychologically safe. As teams prepare to work together, each person should confront their own ability to *experience* feelings of belonging in an inclusive environment (Mather, 2020; Edmondson et al., 2016).

## Inclusion is Dynamic

Inclusion is **dynamic**. Changes to inclusivity occur when workgroups form or end, when the composition of a team changes, or when policies change at the institutional level that impact access to resources or information. While inclusion is informed by an individual's experience of institutional inclusion (which can be more difficult to change), the point of control for the majority of people will be at the *workgroup* level (a level that is much easier to change). Psychological safety in the organization provides the backdrop for experiencing psychological safety within a team or workgroup.

## Inclusion is Ineffable

Because the word "inclusion" refers to the condition of access and the feelings that result from that access, complex meanings and high hopes are often attributed to inclusive environments. By exploring the relationship between inclusion and exclusion, the roles of psychological safety and belonging, the conditionality of inclusion, and the unreasonableness of expecting that inclusive environments are utopian, we come closer to making inclusion more expressible.

**Inclusion and exclusion are interconnected**

Inclusion can not exist without some degree of exclusion.There is no practical organizational design where everyone can (or should) have access to all resources, all information, and all social connections. Some exclusion is not only unavoidable, but desirable. For example, providing access to all resources would quickly become prohibitively expensive, and including everyone in every decision would quickly cause operations to grind to a halt. (Mor Barak et al., 2001) For inclusion to have meaning, the nature of inclusion has to be defined. For example, imagine that a company conducts strategic planning every fall. Which approach is inclusive?

- **Approach #1:** A month prior to the formal sessions, an announcement is sent to the entire company with ideas, guidance, and questions, and a date is set for participants to submit responses. Company leadership reviews and considers all ideas, and a draft plan is shared with the workforce for comment. Finally, the official plan is approved for the upcoming year.
- **Approach #2:** Senior leaders go on a retreat for three days, and two weeks after they return, the organization's goals and plans for the upcoming year are emailed to all employees.

In the first example, each member of the workforce is included in strategic planning through processes that enable their ideas and insights to influence decisions. However, neither of the examples is more likely to lead to a feeling of belonging, or a feeling that your "voice is heard."

**Inclusion benefits from psychological safety and yields feelings of belonging**

> *Diversity is a fact.*
> *Equity is a choice.*
> *Inclusion is an action.*
> *Belonging is an outcome.*
> *-- Arthur Chan, DEI Specialist*

Although psychological safety makes inclusion more *likely*, it is often incorrectly used as a synonym for inclusion. This is detrimental because it suggests that inclusive environments can be cultivated by focusing on feelings over removing blocks to participation. In fact, it is the *ability to participate* that makes it possible for the feelings to emerge. Although inclusion can lead to feelings of belonging, "feeling heard" contributes more to an individual's experience of inclusion and resistance to burnout. (Kerrissey et al., 2022) While inclusion is objective, psychological safety is not:

> *"Even if the 'perfect' organisational mechanisms or team were in place to create Psychological Safety… [it] is a subjective concept that is determined by an individual's beliefs or perceptions of their environment… Despite working in the same environment, employees may attach different meanings to events and interactions depending on their individual state and trait-like characteristics, past experiences and their available psychological resources." (Mather, 2020)*

Distinguishing objective and subjective aspects can help organizations achieve inclusion more quickly, because people "socially construct their perceptions and attitudes based on the social cues within the workplace that in turn influences their behavior" (Boekhorst, 2015) It is insufficient to expect that inclusive behavior emerges automatically when authentic leaders model it, especially in cognitively diverse work environments (with autistic, dyslexic, ADHD, or visually/hearing impaired colleagues). This suggests that teams can benefit from quickly reinforcing desired inclusive behaviors to remove the ambiguity around what enhances safety and what does not.

**Inclusion is conditional**

Inclusion in a group does not necessarily establish inclusion for all time. Once boundaries are defined regarding what constitutes inclusion, community guidelines, rules of engagement, or other social contracts define what is acceptable. Lack of adherence to the contract can result in exclusion.

In most cases there will be some ambiguity around what constitutes a breach of the social contract. Lazerson (2022) calls out four places to strengthen social contracts to protect inclusion even in social virtual worlds. This includes *clarity* regarding when and where community standards apply, *comprehensiveness* of policies to reduce harm, *specificity* about what constitutes a violation, and *transparency* around why standards are in place. Improving expectation setting around inclusion can help protect against the disappointment of inclusive environments not living up to a participant's hopes.

**Inclusion is not conflict-free**

Although inclusive practices can lead to feelings of belonging and "being heard," an inclusive environment is rarely a conflict-free social utopia. Conflict can arise when an employee's perceived level of influence does not align with the actual influence of their role, or when action is inhibited: "there is an unresolved tension over the extent to which voice needs to be effective in changing something in order to be considered voice." (Budd, 2020)

Ferdman (2017) identifies paradoxes of inclusion that embed conflict and contain remedies: self-expression vs. identity, boundaries vs. norms, and safety vs. comfort. First, while authenticity is desirable, expression may be bounded by group identity. Second, adhering to "basic principles of dialogue and democracy" can protect from misaligned expectations about the degree to which each individual's input and consent will be incorporated before actions are taken. Finally, establishing processes for self-determination and power sharing, focusing on freely expressing opinions and influencing decision making around one's *own* work (ie. method, technique, ordering, pace) can help teams distribute discomfort and move beyond oppression. Defining these approaches in advance removes the burden on individuals who may feel powerless to more proactively engage. (McNutt, 2013)

While inclusion can *lead to* feelings of belonging, lack of those feelings does not mean an environment is not inclusive. Inclusion is primarily about ensuring access and removing barriers to access that prevent individuals from achieving deeper understanding or removing objectionable situations. (Allen, 2020)

# Quality and Inclusion

Quality management has long recognized the importance of authentic involvement in one's work, effective teaming, and empowerment to effect meaningful change. By cultivating space for the different backgrounds and ideas people bring, "people start to listen to each other and, once they do, amazing things happen quickly." (Hutson & Perry, 1992) Working transparently and deliberately, and using simple language, contribute to building shared understanding and shared purpose.

### Driving Out Fear

By the 1980s, W. Edwards Deming had documented thinking and practice about individual and team empowerment during his post-war work in Japan, which fueled the Total Quality Management (TQM) movement. (Deming, 1986) Deming's *System of Profound Knowledge* includes *knowledge of psychology* as a pillar, reflecting that human nature plays an important role in successfully achieving quality and performance goals, alongside systems thinking, knowledge of variation, and theory of knowledge. Deming's 14 points call on leaders to break down barriers between departments and silos, and to actively "drive out fear."

When fear is present, honesty suffers. People do not seek out or share truthful information, afraid of negative social or material consequences. This fear has real physiological effects, increasing the heart rate, tensing the muscles, and eroding well-being when sustained over time. "Fear might manifest as blowing up, slamming doors and yelling, or not showing up to a meeting, missing work or even quitting the job. Fear can look like saying nothing in stunned silence, or waiting to see what others will say before answering, and going along with the group rather than sharing a novel idea." (Smith, 2021)

The way to drive out fear is to cultivate psychological safety (one's experience within a team or group). It is the "shared belief held by members of a team that the team is safe for interpersonal risk taking" that allows collaborative learning to occur. (Edmondson, 1999) While psychological safety has been studied at the organizational, team, and one-on-one levels, it is most meaningful and effective at the team level. (Newman et al., 2017) The inclusiveness of a team's leader, the openness and transparency of work within a team, commitment to an improvement orientation, and trusted relationships within the team all contribute to psychological safety. It is different from trust, an interpersonal dynamic.

Lechner and Mortlock (2022) extended this research to virtual teams. They recommend a Plan-Do-Check-Act (PDCA) style approach, leveraging dominance as an asset, engaging in collective goal-setting, and establishing regular checkpoints to assess adherence to the team's self-selected ground rules. Regarding dominance, they recommend that teams identify who they will trust and rely on for technical expertise as part of the planning process. By acknowledging differential skills, teams can channel sources of conflict into improvement. By habitually revisiting the rules of engagement, they provide ongoing opportunities to adjust practices that increase psychological safety.

## Affirming Personal Agency

There is an essential relationship between psychological safety and personal agency. While each person can be affected by those around them, and the work environment, an individual is ultimately responsible for their own reactions, responses, and choices. Reflecting, as a team, on how personal agency impacts an individual's experience of inclusivity can help remove limiting assumptions that would later undermine team performance. This is explained by Sandra Wells, Senior Advisor (Storied, Inc.):

> "Where does psychological safety come from? The idea that someone can make us 'feel unsafe' is erroneous… if you can 'make me feel unsafe' then you're the villain and I'm the victim, and I have no agency regarding my safety. **The feelings I have in the face of your actions will be based on my history of conditioning and my relationship with reactivity.** It's critical that we stop outsourcing our safety to others, and as much as possible, risk exploring the outer edges of our comfort zone. This includes having conversations that are vulnerable, courageous, potentially messy, and very real... and that we each develop the capacity to source approval, validation, and safety from within… [and] engage in mutual learning by being forthright and candid about our own reactivity, defensiveness, or ways we are inclined to withdraw in the face of others' actions."

It is important to recall that "safety" does not mean "freedom from discomfort" and does not instantly yield "feelings of belonging." It is incumbent upon individuals to exercise personal agency and take responsibility for their psychological safety, which in this context means taking the risk to authentically share their thoughts and feelings, accepting whatever discomfort may result, and availing themselves of access to mediation and/or procedural justice processes, if necessary. A key metric could be the frequency over time of instances when such mediation is employed.

## Enabling Decision Making

Team learning, empowerment, and psychological safety have remained core elements of Total Quality Management (TQM) and quality management in general. The inclusion of front-line workers in decision making, attention to openness and transparency to enhance collaboration and cohesiveness, and conscientiously incorporating a variety of perspectives continue to be associated with improved team effectiveness. (Elhuni & Ahmad, 2014)

There is a distinct interplay between inclusion and decision making. Ability to participate in decision making is mentioned explicitly in Tan's (2022) definition, and inclusive decision making contributes to procedural justice. However, concrete guidance on inclusive decision making is limited. Differences of opinion signal opportunities for learning and alignment. If colleagues disagree, it's likely that one has more or different information than the other. It will often be desirable to openly share that information so more complete decisions can be made. In practice, this can be constrained by time and energy available for consensus, or simply by the capacity of the parties to engage.

Figuring out who gets to make decisions (and when) is critical for workgroup inclusion. Teams need concrete guidance on how to establish procedures that maximize the likelihood that 1) everyone feels heard and included, 2) decisions are made in a reasonable amount of time, and 3) good (high quality) decisions are made.

Given the complexity of this aspect of inclusion, the absence of more explicit guidance on establishing decision making processes is intentional. The "right" decision is not always the popular one, and one person's "right" is not necessarily shared. Anticipating decision making protocols can minimize the negative impacts of team members feeling left out, excluded, unheard, ignored, or unjustly overruled. At the same time, teams must continually create a culture where members feel safe to speak their minds, or express unpopular (or simply different) opinions, without fear of negative consequences to their jobs.

# Case Studies

Many organizations prioritize inclusion, but degrees of success vary. While social enterprises prioritize people over profit, traditional organizations may value inclusion as a key aspect of their mission, or aim to capture the increased benefits and reduced costs when employees or participants feel that they have a voice. This section discusses three examples of "successful" inclusion: Zappos, the 541 Eatery & Exchange, and Burning Man. In each of these cases, the majority of participants agree that an inclusive environment is present, and provide insights into the costs of making it happen.

**Holacracy at Zappos**

The management philosophy of *holacracy* aims to fully democratize organizational design by eliminating the concept of management. Holacracy gives teams and workgroups complete power to self-organize and establish boundaries with other groups. While relatively rare, there is one high profile example that demonstrates how performance and inclusion may be interrelated: shoe company Zappos, founded by Tony Hsieh in 2000, which transitioned to holacracy between 2012 and 2015 by removing the hierarchical reporting system and investing heavily in holacracy training.

Also referred to as radical decentralization, holacracy shifts the locus of decision making from managers to individuals, and purports to empower anyone to have a voice. It requires the explicit definition of roles and accountabilities. Theoretically, when people know exactly what it is they are responsible for (and have complete authority to do it) a company should see increased employee motivation, satisfaction, and productivity.

In reality, results are mixed. Street & Feeney (2022) explain that while increased satisfaction due to "having a voice" is indeed a theme, "efficacy… is influenced by whether employees *believe* that everyone, including those in senior roles, are genuine in their intent and are following the same rules." The paradox extends to work practices. One respondent "liked that the new system gave everyone a voice in meetings… [but] created very long meetings."

While inclusion and belonging increased with holacracy, stronger information silos tended to form more quickly. Several of Zappos leaders were frustrated by the sluggishness of progress. "Like in a democracy, decisions get made really, really slowly. You can't *make* anyone do anything, and lots of people take advantage of that. Sure I feel included in all discussions, but it's nearly impossible to make progress. I think the business suffers because of it." (Radziwill, 2013)

The Zappos case provides two key insights about how to cultivate inclusion through empowerment. First, focus "away from the personal and instead on the work itself." (Robertson, 2015) Emphasizing the access aspects of work design can create the conditions for the personal feelings to emerge. Second, participants acknowledged that personal beliefs about fairness or justice might influence an individual's ability to meaningfully participate in an inclusive environment, but no approaches for making this happen were identified.

## Inclusion at 541

The 541 Eatery & Exchange in Hamilton, Ontario is a non-profit restaurant with a mission to serve the community through a "pay it forward" approach. Although the company has some full time staff, most of the workforce consists of volunteers. Many are young or disabled, and seek the opportunity to gain job skills.

Scott & Wilton (2021) performed an extensive case study of 541, which is "uniquely positioned to partly redress disabled people's exclusion from the workplace." 541 designed many inclusive day-to-day practices, starting with a low pressure environment where workers are not required to achieve strict standards of performance. Scheduling flexibility is the norm (due to increased levels of fatigue). Experimentation, creativity, and adaptation are encouraged to help the job fit into the context of each worker's life. Regular accessibility audits ensure physical inclusion. Managers are trained to carry out their "dual roles" of emotional labor and restaurant operations, and gently correct errors where needed. Most starkly, 541 does not "adhere to ableist standards of productivity" and acknowledges that service may be slow or not to the standards of other restaurants.

Although many workers report a strong experience of inclusion, and feelings of belonging, inclusive practices have not made this a guarantee. "Assumptions that… social enterprises designed for disabled people… or inclusive spaces… can consistently engender feelings of belonging for all disabled people across time has also been called into question." (Scott & Wilton, 2021) Managers report that inclusion takes an immense amount of energy to sustain. They question whether they are being too supportive or flexible, or whether focusing on inclusion over efficiency is feasible because of the stress it puts on the business. Though many are exhausted by the emotional labor, "the staff exhibit a deep commitment to the restaurant's mandate despite the toll it takes on individuals."

In all cases, 541 chooses people over performance in order to further its mission, and is able to do this because it relies heavily on volunteers. Not all social enterprises have this ability. To survive, they must "make difficult decisions about how to best balance their social mandate against the reality of a competitive marketplace… [which will] directly inform the nature of the work environment and experiences of their disabled workers." (Scott & Wilton, 2021)

## Radical Inclusion

The most extreme example of an environment that is inclusive by design is the annual Burning Man event, which takes place the week before labor day in the Black Rock Desert, a hundred miles away from the nearest city. Attended by nearly 80,000, Black Rock City is the third largest city in Nevada after it emerges from nothing. It disappears into the nothingness of the harsh, remote desert a couple weeks later.

Unlike festivals, Burning Man has many elements you might see in any city: a fully functioning airport sanctioned by the FAA, a fully operable post office, cafes, bike shops, dance clubs, and at least one roller rink. These are all contributed by participants and "gifted" to Burning Man's citizens. There is a tightly-knit group of peacekeepers and mediators called the Black Rock Rangers, who mediate disputes and serve as liaisons to local and federal law enforcement, and the Department of Public Works (DPW) manages the city's infrastructure. Garbage cans are absent: one of the ten principles of the city's social contract is "Leave No Trace" and each person is responsible to ensure there is no "matter out of place" (MOOP). The other principles are radical inclusion, gifting, decommodification, radical self-reliance, radical self-expression, communal effort, civic responsibility, participation, and immediacy.

"Radical Inclusion" means *anyone* can participate, *everyone* is welcome and respected, and there are *no prerequisites* for participation. A desire to participate in the community, and the willingness to adhere to its social contract, is all that is needed. "Burners do not attend Burning Man as spectators or consumers, but seeking to be incorporated into the social fabric… expected to help build their temporary metropolis, to keep it running, and to help maintain the social equilibrium." (Gomez 2013) The social order is maintained primarily by the participants, who are "willing to cooperate voluntarily" and committed to creating an environment *for each other* that values inclusion, self-expression, and authenticity.

Despite these shared values, there are still problems. Boundary issues, improper use of personal property (consent), and "improper behavior" (typically, overstaying your welcome at someone else's camp) are frequently mediated by the Rangers, and occasionally outside law enforcement removes people who are threatening others. Halcyon (2011) explains that "radical inclusion does not mean that every door is open all the time… [you can't] take whatever you want, it doesn't mean that you can join [a group] automatically. The default is, the door is open. But if you see a closed door, you have to knock. Radical inclusion is the big picture... [but you're] responsible for crafting your experience."

Radical inclusion is challenging because "putting up with our radical differences takes work." (Perez-Banuet, 2014). Although many participants report a strong experience of inclusion and feelings of belonging, it is still possible to feel lonely and lack a sense of belonging. Many agree. Wachs (2015) explains that radical inclusion feels like work: "You have to let people play with you in a meaningful way… if it's easy to include someone, that's not radical inclusion, that's self interest. It's not radical inclusion if it's exactly what you would have done anyway, if it doesn't make you uncomfortable, if it doesn't expand the people you play with. It doesn't mean being nice to people you were already going to be nice to. It's giving up control… putting aside the power to say 'people like you are in my world, and people like you are not'. You don't have to approve of everything anyone does, but you do need to give people the opportunity to participate in a meaningful way."

From the case of Burning Man, several lessons regarding successful inclusion emerge: a simple, clear social contract; well-defined decision making processes coupled with the availability of mediation; free choice behaviors instead of behavioral compliance; lack of differentiated status; and the assumption of self-reliance. (Hoover et al., 2017) Inclusion is an *agreement* to help each other find meaningful ways to contribute, and to expand comfort zones to introduce new opportunities for relating.

# Discussion & Conclusions

The Engagement-Performance Paradox refers to the potential conflict between fostering an inclusive, engaging work environment and maintaining high levels of performance and productivity.This study identified elements to help teams become more inclusive by design, managing this potential tension to reap the benefits of inclusion without sacrificing performance. A literature review of research spanning inclusivity and performance, combined with lessons from three case studies, revealed that inclusion by design requires:

**Plan:**
1. Engaging in intensive expectation setting about *what inclusion is and is not* when teams are created (or composition changes);
2. Recognizing that each individual joins a team with past experiences, traumas, and biases that might influence their *personal experience* of team inclusion, including their experience of organizational justice at the institution that the team is part of;
3. Helping individuals become aware of the *value of self-reliance* in affirming agency and resisting the perception of victimhood for developing feelings of psychological safety;

**Do:**
4. Guiding teams to collaboratively and non-prescriptively *co-create an inclusive environment*, to remove the assumption that inclusion is "provided" by a dominant group;
5. Constructing *working agreements* that define how team members will create an inclusive environment for each other, leveraging mediation when necessary, and maintaining standards for quality and performance;
6. Defining lean *decision making processes* in advance, because effective decision making is often differential rather than democratic;

**Check:**
7. Developing an *assessment cadence* where teams can quickly determine how well they are meeting self-determined inclusion objectives;

**Act:**
8. Establishing a habit of *quick incremental action*, when assessment identifies gaps, to achieve steady continuous improvement of psychological safety and work outcomes.

While efforts are beginning to emerge to help teams become more inclusive (see, e.g. Davila et al. 2020), we still need to transform inclusion from an aspirational ideal to a practical, measurable outcome. This work takes the first step by articulating eight steps to help teams become inclusive by design using Plan-Do-Check-Act (PDCA). By framing inclusion as "something teams do" instead of "something teams are given," control is returned to those who are best positioned to deliver on the aspiration.

## Acknowledgements

This work was supported by Ultranauts, Inc. under funding from Neurodiversity in the Workplace (NITW) to help teams become more inclusive by design. This research formed the basis for the http://team-x.ai/ product that helps teams identify the inclusive action that will support the psychological base states of its unique combination of members. Ultranauts is the world's first fully remote workplace for cognitively diverse teams, 75% of whom are neurodivergent (autistic, dyslexic, non-speaking, and/or hard of hearing). The firm partners with large enterprises across a range of industries to improve software and data quality and create a Universal Workplace that embraces diversity and serves as a replicable blueprint for others. The authors report there are no competing interests to declare.